\documentclass[lettersize,journal]{IEEEtran}

\usepackage{cite}
\usepackage{amsmath,amssymb,amsfonts}
\usepackage{algorithmic}
\usepackage{graphicx}
\usepackage{textcomp}
\usepackage[table]{xcolor}
\usepackage{booktabs}
\usepackage{enumitem}
\usepackage{tabularx}
\usepackage{amsmath}
\usepackage{graphicx}
\usepackage{url}
\usepackage{listings}
\usepackage{tikz}
\usepackage{tcolorbox}

\usepackage{booktabs}
\usepackage{tabularx}
\usepackage{array}

\usepackage{multirow}
\usepackage{makecell}
\usepackage{amssymb}
\usepackage[hidelinks]{hyperref}
\newcommand{\todo}[1]{}

\definecolor{herpurple}{RGB}{245,240,255}

\newcommand{\artifactrepo}{\url{https://github.com/BeeuniLab/pheragent/tree/main}}
\lstdefinestyle{cstyle}{
language=c,
basicstyle=\ttfamily\bfseries\small,
  morekeywords={virtualinvoke},
  keywordstyle=\color{black},
  ndkeywordstyle=\color{red},
  commentstyle=\color{dkred},
  stringstyle=\color{ForestGreen},
  numbers=left,
  breaklines=true,
  numberstyle=\ttfamily\footnotesize\color{gray},
  stepnumber=1,
  numbersep=10pt,
  backgroundcolor=\color{white},
  tabsize=4,
  showspaces=false,
  showstringspaces=false,
  xleftmargin=.23in,
  captionpos=b,
  escapeinside={\$}{\$}
}

\newcommand\lt[1]{{\lstinline!#1!}}

\def\BibTeX{{\rm B\kern-.05em{\sc i\kern-.025em b}\kern-.08em
    T\kern-.1667em\lower.7ex\hbox{E}\kern-.125emX}}
\begin{document}

\title{HerAgent: A Block-Level Checkpointing Harness for Coding-Agent Environment Setup}

\author{Xiang~Li, Siyu~Lu, Federica~Sarro, Claire~Le~Goues, and He~Ye%
\thanks{Xiang Li, Federica Sarro, and He Ye are with University College London, United Kingdom.}%
\thanks{Siyu Lu is with Uppsala University, Sweden.}%
\thanks{Claire Le Goues is with Carnegie Mellon University, United States.}}

\markboth{IEEE Transactions on Software Engineering,~Vol.~XX, No.~X, Month~Year}%
{Li et al.: HerAgent}

\maketitle
\begin{abstract}

Modern coding agents rely on establishing executable software environments before they can perform software engineering tasks such as testing, debugging, and patch validation.
Existing setup agents typically control progress at either single-command or whole-script granularity, making recovery either too local to remove polluted state or too coarse to preserve useful progress. 
We present HerAgent (\underline{H}arness \underline{E}nvi\underline{R}onment Setup Agent), a block-level harness for coding-agent environment setup. HerAgent's main contribution is semantic setup blocks: a single progress-control abstraction used for execution, validation, checkpointing, repair, and rollback.
{Evaluated against six prior agents on four environment-setup datasets, HerAgent improves setup success in every comparison, achieving the largest gains on Installamatic-Bench (52.0\% to 85.0\%) and SetupBench (46.3\% to 70.4\%). It also reduces setup costs by up to 94.5\%, showing that block-level progress control improves reliability and efficiency.}
\end{abstract}



\section{Introduction}
Modern software engineering agents are increasingly expected to operate on real repositories, carrying out tests that depend on a working build and runtime~\cite{bouzenia2025repairagent,xia2024agentless,bouzenia2025understanding, yang2024sweagent, jimenez2024swebench}. 
To succeed, agents must first reconstruct system packages~\cite{arora2025setupbench}, language runtimes~\cite{eliseeva2025envbench,huang2026rat}, project-level dependencies~\cite{latendresse2022not,hu2025repo2run}, and build or CI assumptions~\cite{santolucito2018statically,rausch2017empirical}.
This process is long-horizon, stateful, and failure-prone~\cite{rahman2020gang, rahman2018bugs, gupta2024continuous, rahman2018characterizing}.
Automated environment setup is therefore a prerequisite for practically useful AI coding agents.

Existing work has made important progress on coding-agent environment setup~\cite{arora2025setupbench, bouzenia2025you,hu2025repo2run,kovrigin2025piper,eliseeva2025envbench, guo2026menvagent, milliken2025beyond, guo2025swefactory, wang2024openhands, yang2023auto}. 
At a high level, these systems combine an LLM-driven controller with an execution harness: the agent inspects the repository and proposes setup actions, while the harness executes them in a sandbox and returns observations. Output is typically either a command trajectory or a generated setup artifact (e.g., a Bash script or Dockerfile).

Long setup trajectories often require the agent to validate partial progress and recover from invalid states after failed actions~\cite{bouzenia2025understanding, zhao2026debugging, hutter2026agentstepper}.  
A major reason both validation and recovery are difficult is that the correctness signal provided by an individual setup or install step is local and weak relative to the overall success criterion~\cite{beller2017oops, beller2017travistorrent, gallaba2018use, seo2014programmers}.
For example, one important step in environmental configuration is installing necessary package dependencies~\cite{latendresse2022not, macho2018automatically, wang2018dependency}. However, a package manager can locally only report whether a package installed correctly. It cannot confirm that the version is correct, nor that it will not conflict with other packages installed at different steps (before or after)~\cite{zhou2026setupx}. 
The global check entails evaluating whether the overall project builds and runs; this can only happen at or near the end~\cite{gallaba2018use, beller2017oops, chen2026swe}.
Failures therefore do not always correspond to a problem in the most recent command, but may be the result of an earlier decision~\cite{xu2025openrca, zhao2026debugging}. 
%
Setup agents~\cite{zhao2026debugging} consequently typically track intermediate environment states to support failure recovery (rollback) and retry. 

This checkpoint and recovery mechanism must balance effective failure recovery with a clean, replayable setup artifact~\cite{merkel2014docker, dolstra2004nix, chirigati2016reprozip}.
We define \emph{replayability} in this context as the ability to cleanly reconstruct project setup in a fresh environment based only on the output setup artifact (not requiring the live container)~\cite{merkel2014docker, dolstra2004nix, chirigati2016reprozip}. 
Prior work maintains intermediate setup state at different granularity levels~\cite{bouzenia2025you,hu2025repo2run,kovrigin2025piper,milliken2025beyond}, affecting this tradeoff. 
At one end, a \emph{command-level harness} (e.g., ExecutionAgent~\cite{bouzenia2025you} or Repo2Run~\cite{hu2025repo2run}) advances one command at a time, attaching recovery to individual command executions.
However, this approach can struggle to recover from earlier decisions, and a successful setup with recovery is often not cleanly replayable. 
Repo2Run, for example, 
uses per-command Docker snapshots, but its rollback mechanism improves runnability by only 3.1\%~\cite{hu2025repo2run}.
At the opposite extreme, \emph{whole-script harnesses}~\cite{kovrigin2025piper, milliken2025beyond} treat an entire Dockerfile or Bash script as the recovery unit, producing clean trajectories by construction, but making failures hard to localize.  
Installamatic, for example, recovers from failures by rebuilding with a fresh agent, capping repair at two attempts. This avoids polluted state but discards intermediate progress~\cite{milliken2025beyond}.

These extremes expose a \emph{granularity gap} for setup harnesses.  Neither end offers a unit of progress small enough to localize repair, while large enough to support clean replay. 

In this paper, we propose \textbf{HerAgent} (\textbf{H}arness \textbf{E}nvi\textbf{R}onment Setup Agent), a coding-agent environment-setup system that maintains progress as an ordered sequence of \emph{block-level} \texttt{BashFile}s and sandbox checkpoints.
Our key insight is that environment setup proceeds through discrete functional stages, such as installing system dependencies, provisioning a language runtime, or resolving project-specific dependencies.
Each stage groups related commands and ends at a natural validation boundary marking a known-good intermediate restart point. 
A \emph{block} therefore groups the dependent commands of one functional stage, and serves as the unit for planning, validation, checkpointing, and recovery. We call the resulting mechanism a \emph{Block-Level Checkpoint}.

HerAgent begins by inspecting repository artifacts, and then partitions expected setup actions into functional blocks. 
It advances block by block with validation, and checkpoints each validated state. 
On failure, it attempts bounded local repair, accepting a fix only if the patched block replays from the previous validated checkpoint. 
If the evidence indicates that an earlier block introduced the faulty state, HerAgent rolls back to that block's preceding checkpoint and resumes recovery there.
Once all blocks validate, HerAgent replays the finalized sequence in a clean container, making accepted setups replayable by construction.

To our knowledge, HerAgent is the first harness to track and recover coding-agent setup state at functional-block granularity. {Across four datasets and six prior agents, it improves setup success in the reported matched-model comparisons; its largest cost reduction is 94.5\% relative to OpenHands on SetupBench using GPT-4.1.} HerAgent reduces wasted exploration, preserves validated progress, and reconstructs environments from a clean state.
We summarize our contributions as follows:
\begin{description}[itemsep=0pt, topsep=3pt, parsep=0pt, font=\normalfont\bfseries]
    \item[Problem.] We formulate coding-agent environment setup as a \emph{progress-control} problem and identify granularity as a key limitation of existing harnesses.
    \item[Algorithm.] We introduce \emph{Block-Level Checkpointing}, where setup advances over validated functional blocks and repairs must replay from clean checkpoints.
    \item[Framework.] We design and implement HerAgent, a novel setup harness that combines block-level validation, local repair, and root-cause rollback.
    \item[Evaluation.] We evaluate HerAgent against six prior agents on four datasets, outperforming all baselines.
    \item[Open Science.] We release our implementation, benchmark subset, validation commands, and experiment artifacts.\footnote{\artifactrepo}
\end{description}

\section{Motivation}

\begin{figure}[tb]
\centering
\includegraphics[width=0.75\linewidth]{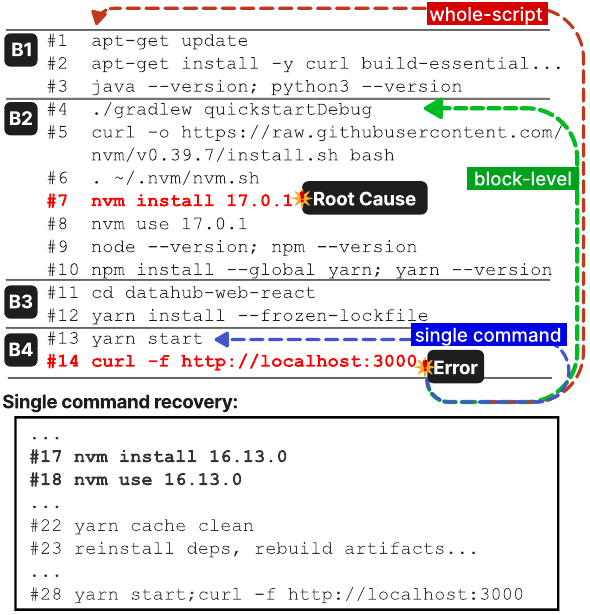}
\caption{A portion of a command trajectory in configuring DataHub. An agent successfully installs the wrong Node version at \textbf{\#7}. The frontend validation at \textbf{\#14} exposes the failure.  Whole-script recovery restarts from scratch; single-command recovery starts from the failed command and must clean polluted state; block-based recovery rolls back to the block introducing the faulty state.}
\label{fig:motivating-example}
\end{figure}

Figure~\ref{fig:motivating-example} shows a setup attempt for DataHub, which combines Java services, Python tooling, and a Node/Yarn frontend.\footnote{\url{https://github.com/datahub-project/datahub}; setup commands discussed in the developer guide at \url{https://docs.datahub.com/docs/developers}.}
The agent successfully installs Node 17 (\#7), but this version is wrong because it ships with OpenSSL 3, which removes the legacy hashing required by the frontend's WebPack version. The failure surfaces only later during frontend validation (\#14).



\emph{Why does single-command reasoning and recovery fail?}
A single-command harness (like ExecutionAgent~\cite{bouzenia2025you} or Repo2Run's command-level execution and rollback setting~\cite{hu2025repo2run}), advances one command at a time, asking the model to reason about each separately. 
Per-command reasoning is often wasteful, however, because many commands are semantically coupled.  For example, 
once the system decides to provision the runtime, launching the local stack (installing \texttt{nvm}, selecting Node, checking versions, installing Yarn) can be generated and validated as one functional unit.
Additionally, for recovery, rolling back to only command \#13 leaves the incorrect runtime and dependency state intact. This requires the agent to engage in cleanup and repeated local trials, shown on the blue single-command recovery path (\#17--\#28)

\emph{Why does whole-script reasoning and recovery fail?} 
The opposite extreme is whole-script generation, like in PiPer~\cite{kovrigin2025piper} or Installamatic~\cite{milliken2025beyond}, avoids per-command reasoning and generates a complete Bash script or Dockerfile at once.
However, this means failed runs can require regenerating and rerunning a large artifact (the red whole-script recovery path in Figure~\ref{fig:motivating-example}). This wastes useful progress, like the successful earlier system and language dependency setup here (\#1--\#3).
Repeated whole-script regeneration can also introduce new hallucinated changes into previously correct commands, compromising stages that had previously succeeded.



These limitations motivate an intermediate unit of progress control. We group related setup commands into replayable \emph{blocks}, denoted $B_1, B_2, \ldots$. Each block is a semantically coherent group of dependent commands that implements one functional setup stage. 
In Figure~\ref{fig:motivating-example}, the DataHub trajectory naturally falls into four blocks. $B_1$ installs system and language dependencies (\#1--\#3); $B_2$ starts the local stack and provisions Node/Yarn (\#4--\#10); $B_3$ installs frontend dependencies (\#11--\#12); $B_4$ validates frontend startup (\#13--\#14). HerAgent therefore rolls back to the checkpoint before $B_2$, where the incorrect node version was introduced, repairs it, and replays $B_2$--$B_4$. This preserves the validated prefix $B_1$, avoiding both long cleanup and whole-script regeneration.

\section{Approach}

\begin{figure*}[tb]
\centering
\includegraphics[width=0.65\linewidth]{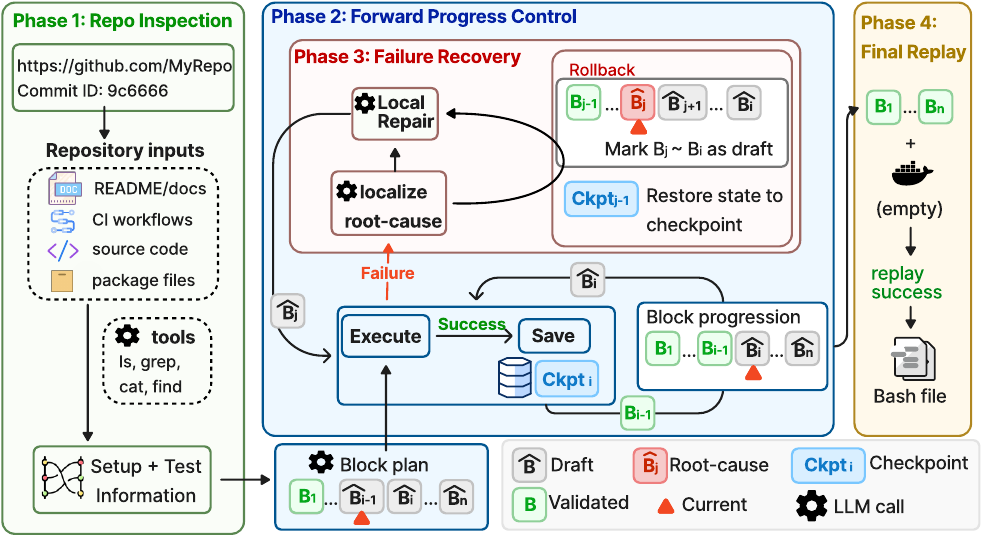}
\caption{Overview of HerAgent. Phase 1, Repository inspection collects setup and validation evidence. Phase 2, Forward progress control executes, validates, and checkpoints setup blocks. Phase 3, Failure recovery repairs failed blocks or rolls back to earlier checkpoints. Phase 4, Final replay verifies the validated block sequence from a clean environment to produce a reproducible setup artifact.}
\label{fig:approach}
\end{figure*}

This Section presents HerAgent, which automatically configures a repository's environment at the granularity of functional blocks, the boundaries at which it  validates progress, checkpoints state, and localizes recovery.
As shown in \autoref{fig:approach}, HerAgent takes a repository URL, a tag or commit, and an initial Dockerfile, then runs four phases. (1) \textbf{Repository Inspection} (Section~\ref{sec:inspection}, green in \autoref{fig:approach}) builds an isolated container, clones the repository, and collects setup and validation evidence; (2) \textbf{Forward Progress Control} (Section~\ref{sec:forward}, blue) generates setup blocks and advances them through execution, validation, and checkpointing; (3) \textbf{Failure Recovery} (Section~\ref{sec:recovery}, red) repairs failed blocks locally, patches accepted fixes back, and relocates recovery to an earlier checkpoint when needed; and (4) \textbf{Final Replay} (Section~\ref{sec:replay}, yellow) reruns the finalized block sequence from a clean environment and outputs the executable blocks, validated final container, and validation report.

\subsection{Repository Inspection}
\label{sec:inspection}

HerAgent first collects repository evidence as context (Phase~1 in \autoref{fig:approach}) before generating or executing setup blocks. Using only non-destructive read-only Bash commands (e.g., \texttt{ls}, \texttt{find}, \texttt{grep}, and \texttt{cat}), the agent navigates the repository to extract two types of information:
(1) \emph{setup evidence}, identifying detected language ecosystems (e.g., Java, Python, or both) and build files (e.g., \texttt{requirements.txt}), to construct the "Bash Script" part of each block, and (2) \emph{test evidence} via identified testing directories and test frameworks (e.g., \texttt{pytest} and \texttt{mvn test}). { This evidence is used to construct the $V$ part of each block.}
These artifacts become prompt context for block construction in the next phase. 

\subsection{Forward Progress Control}
\label{sec:forward}
During forward progress control (Phase~2 in \autoref{fig:approach}), HerAgent builds environment setups by constructing blocks, controlling progress, and checkpointing.

\subsubsection{\textbf{Constructing Blocks}}
HerAgent constructs environment setup as an ordered sequence of executable blocks. Before executing setup commands, HerAgent uses the repository evidence to synthesize the complete block sequence
\[
\begin{aligned}
\mathcal{B} &= [B_1,\ldots,B_n].
\end{aligned}
\]
Each block is represented as
\[
\begin{aligned}
B_i = \langle Hash, Name, Purpose, \\
Prelude, Bash Script, V \rangle .
\end{aligned}
\]
\emph{Name} and \emph{Purpose} summarize the functional stage and intended environment state;  \emph{Hash} is a unique per-block identifier. 
The \emph{Prelude} and \emph{Bash Script} are generated together during initial block construction. 
\emph{Prelude} stores persistent shell context, such as virtual-environment locations and working-directory bindings, inherited from earlier blocks and extended.
The \emph{Bash Script} is the executable  commands, written for non-interactive \texttt{POSIX} \texttt{sh}, supporting portability. 
{ $V$ stores the associated validation targets, which are lightweight checks for the produced intermediate state.
The LLM generates each block and its corresponding $V$ from the setup and test evidence collected during Repository Inspection, tailoring the checks to the block's functional purpose. For example, a test-tooling block may use \texttt{pytest --collect-only} to check whether tests can be discovered. Blocks are emitted in JSON, with a non-interactive \texttt{POSIX sh} validation command for each block. These checks assess the state produced by the current stage; they do not replace the benchmark's external oracle for final task success. A block is checkpointed only when both its setup commands and its validation succeed.}

Repository setup varies, but many stages recur at a high level.
HerAgent uses seven block categories (Table~\ref{tab:block-templates}) as LLM guidance to capture recurring responsibilities, dependency order, and semantic rather than arbitrary block boundaries.
Before execution begins, HerAgent partitions the setup procedure into an ordered sequence of blocks according to function, such as installing system packages, provisioning language runtimes, installing project dependencies, configuring native builds, or preparing test tooling. Commands that serve the same functional goal and share a validation target are grouped into one block.
The evidence collected during Repository Inspection guides the initial selection of relevant categories: single-language projects typically use runtime, dependency, and test/tooling blocks, while multi-language or compiled projects may add additional-runtime, additional-dependency, or native-build blocks.
During this initial construction, categories may be merged, split, or omitted as needed to keep each block executable and locally checkable.




 \begin{table}[tb]
    \caption{Categories of block templates used by HerAgent.
    }
    \label{tab:block-templates}
    \centering
    \footnotesize
    \setlength{\tabcolsep}{4pt}
    \renewcommand{\arraystretch}{1.02}
    \begin{tabularx}{\columnwidth}{@{}p{0.30\columnwidth}|>{\raggedright\arraybackslash}X@{}}
    \toprule
    \textbf{Block template} & \textbf{Function} \\
    \midrule
    \emph{System packages}
    & Install shared OS prerequisites, e.g., \texttt{apt-get install git curl}. \\
    \midrule
    \emph{Language runtime}
    & Provision or verify the primary runtime, e.g., \texttt{python3 -m venv .venv}. \\
    \midrule
    \emph{Project dependencies}
    & Install the primary ecosystem's dependencies, e.g., \texttt{pip install -r requirements.txt}. \\
    \midrule
    \emph{Additional / merged runtime}
    & Provision an additional runtime or merge multiple runtimes, e.g., \texttt{node --version}. \\
    \midrule
    \emph{Additional / merged dependencies}
    & Install additional dependencies or merge multiple ecosystems, e.g., \texttt{npm install}. \\
    \midrule
    \emph{Native build config}
    & Verify compilers and native build tools, e.g., \texttt{gcc --version \&\& cmake --version}. \\
    \midrule
    \emph{Test tooling}
    & Prepare lightweight validation checks, e.g., \texttt{pytest --collect-only -q}. \\
    \bottomrule
    \end{tabularx}
    \end{table}



\subsubsection{\textbf{Constructing Checkpoints}}
HerAgent advances repository setup by executing and validating drafted blocks in order and records checkpoints after each successfully executed block.
Formally, after validating $B_i$ via $V_i$, HerAgent records
\[
\begin{aligned}
Ckpt_i =
\langle
B_i, D_i, R_i
\rangle , \mathtt{if}\ i > 0;\\
\langle D_i, \rangle , \mathtt{if}\ i == 0.
\end{aligned}
\]
$D_i$ is the Docker state after executing the validated prefix $B_1,\ldots,B_i$.  $R_i$ stores exit codes and execution logs for the $B_i$ that justify the checkpoint, which later repair and rollback decisions can use to determine which checkpoint to restore.
Each checkpoint $Ckpt_i$ is uniquely bound to block $B_i$ by $Hash$.
The implementation uses \lt{docker commit} and restores checkpoints by launching a new container from that image using \lt{docker run}.


\subsubsection{\textbf{Block-level Forward Progress}}
HerAgent tracks forward progress as a sequence $\mathcal{B}$ partitioned into validated $B_1,\ldots,B_k$ and draft blocks $\widehat{B}_{k+1},\ldots,\widehat{B}n$:
\[
\mathcal{B} = [B_1,\ldots,B_k,\widehat{B}{k+1},\ldots,\widehat{B}_n],
\]
($\widehat{B}$ denotes draft).
Checkpoints enforce sequential dependency, and blocks are trusted only if the preceding prefix is verified. If a prior block $B_i$ is modified, HerAgent invalidates all downstream blocks and discards their checkpoints.
To advance a draft block $\widehat{B}_{k+1}$ into a validated state, HerAgent executes \emph{Prelude}, \emph{Bash Script}, and validation $V$. If execution succeeds, the system records execution results $R_{k+1}$, saves the runtime Docker environment as $Ckpt_{k+1}$, and promotes $\widehat{B}_{k+1}$ to $B_{k+1}$. Otherwise, $\widehat{B}_{k+1}$ remains unvalidated and the system enters the Failure Recovery phase.

\vspace{-1ex}
\subsection{Failure Recovery}
\label{sec:recovery}

\begin{figure*}[tb]
\centering
\includegraphics[width=0.8\linewidth]{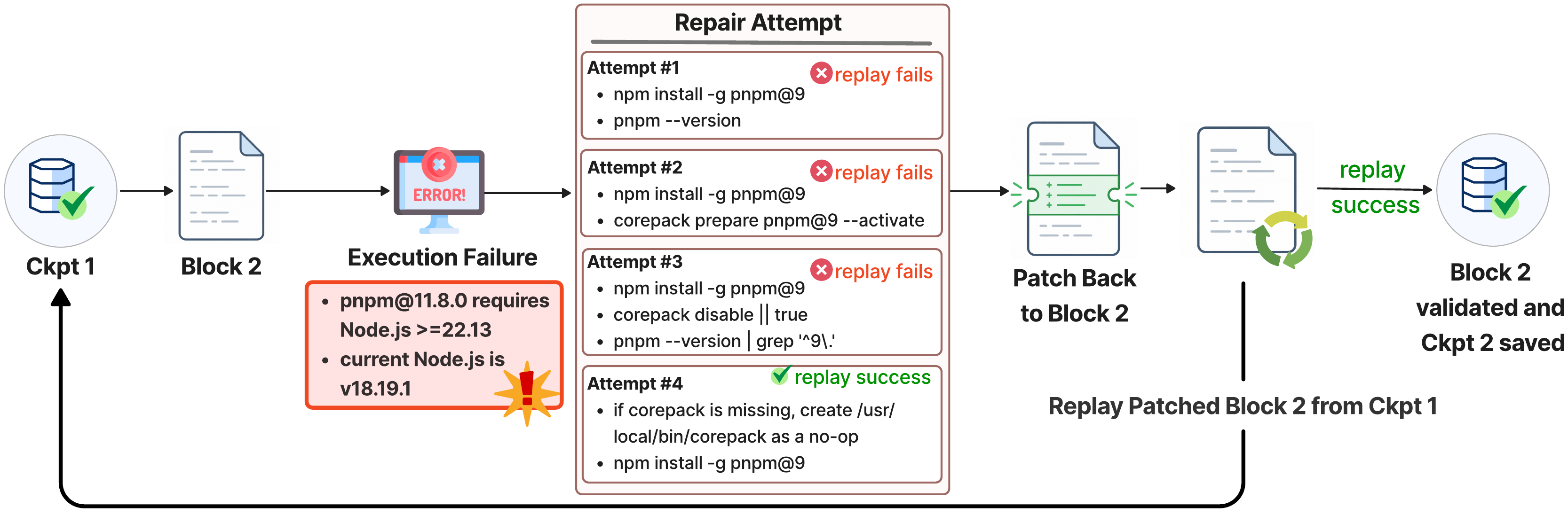}
\caption{Local repair starts from the previous validated checkpoint, tries a bounded sequence of candidate fixes for the active failing block, patches a successful fix back into the block artifact, and accepts the repair only if the patched block replays successfully and the next checkpoint is saved.}
\label{fig:localrepair}
\end{figure*}

If block execution or validation fails in Phase~2, HerAgent enters Failure Recovery (Phase~3 in \autoref{fig:approach}).
It first attempts \emph{local repair} within the failed block $B_i$, incorporating fixes into durable block content for replay.
If evidence points to an earlier faulty block, HerAgent escalates to rollback and failure relocation.

\paragraph{Local Repair}

Figure~\ref{fig:localrepair} shows the local-repair loop for a failing block $B_i$. Starting from the last validated checkpoint $Ckpt_{i-1}$, HerAgent executes $B_i$ to observe the failure. It uses the error log to diagnose a repair target inside $B_i$. A \texttt{pnpm}/Node.js version mismatch, for example, localizes repair to the package-install step inside the current active block.

The loop in Figure~\ref{fig:localrepair} follows three steps: (1) \emph{attempt}, where HerAgent proposes and executes a candidate fix for $B_i$ in the live container; (2) \emph{patch-back}, which merges a successful fix into the block artifact (either the \emph{Bash Script} for executable behavior or the \emph{Prelude} for persistent shell context such as \texttt{PATH} definition); and (3) \emph{replay}, where HerAgent reruns the patched block from a \emph{clean} $Ckpt_{i-1}$. Failed attempts or replays are rejected; HerAgent retries the loop within a bounded budget. If replay succeeds, HerAgent promotes the patched block to a validated state, and saves a fresh $Ckpt_i$.
If no candidate passes within the repair budget, recovery escalates to failure relocation.

\paragraph{Failure Relocation and Checkpoint Rollback}
HerAgent uses failure relocation to reason about failures that are likely caused by blocks prior to $B_i$ (where the symptom appears). 
To do this, the LLM controller reasons about the failing command in $B_i$ along with $R_i$,  block scripts, and checkpoints of the validated prefix $\langle Ckpt_1,\ldots,Ckpt_{i-1}\rangle$. It proposes an earlier root-cause block $B_j$ with $j<i$, invalidates $B_j$ and all later blocks as $\widehat{B}_j, \widehat{B}_{j+1}, \ldots$, restores $Ckpt_{j-1}$, prunes later checkpoints, and resumes local repair from $\widehat{B}_j$. This is the Rollback transition shown in \autoref{fig:approach}.
Rollback avoids both command-level search and full reset by discarding only the suspected faulty suffix. 

\vspace{-1ex}
\subsection{Final Replay}
\label{sec:replay}
During the final replay (Phase~4 in \autoref{fig:approach}), HerAgent replays the validated block sequence in a clean environment. After all blocks are validated, HerAgent reruns the finalized sequence in a clean container, using only the durable block artifact, including each block's \emph{Prelude}, \emph{Bash Script}, and validation $V$. Interactive repair can leave hidden state that makes a setup look successful when it is not, so HerAgent accepts the setup only when this clean rerun also succeeds.

This final replay prevents HerAgent from accepting a setup that only worked because of unstated side effects from live repair, such as transient environment variables, manually changed directories, cached packages, or commands that were executed interactively but never patched back into the artifact. If the replay succeeds, HerAgent outputs the reproducible setup artifact and its final validation report. If it fails, HerAgent reports a reproducibility failure and attributes it to the corresponding block or validation phase.


\section{Experiments}

\label{sec:experiments}

We evaluate HerAgent by measuring performance, ablating components, analyzing recovery trajectories, and examining failures. Thus, we ask four research questions:

\textbf{RQ1: Overall Effectiveness.} \emph{How well does HerAgent perform on existing setup benchmarks?}

\textbf{RQ2: Component Contribution.} \emph{Which components of HerAgent contribute to its effectiveness?}

\textbf{RQ3: Trajectory-Level Recovery Analysis.} \emph{How does HerAgent structure setup into blocks (\emph{RQ3.a}) and use repair and rollback across trajectories (\emph{RQ3.b})?}

\textbf{RQ4: Failure Analysis.} What causes HerAgent's failures?

\subsection{Datasets and Baselines}

\paragraph{Datasets}
We evaluate on four diverse environment-setup benchmarks:

\noindent\textbf{ExecutionAgent-Bench}~\cite{bouzenia2025you} consists of 50 repositories across 14 programming languages with test commands derived from CI workflows. A run succeeds when the build and target test suite execute, matching expected pass, fail, and skip counts. 

\noindent\textbf{SetupBench}~\cite{arora2025setupbench} contains 93
  environment-bootstrap tasks requiring package installation, dependency
  resolution, database initialization, and background-service configuration in a
  fresh Linux sandbox; we use its Repo Setup subset of 54 repositories across seven ecosystems (Python, TypeScript, JavaScript, Go, Rust, Java, and C++). It
  checks repository readiness through user-intended success commands rather than
  only test-suite execution, and a run succeeds when the benchmark success command  prints \lt{Setup successful}. 

\noindent\textbf{Repo2Run-Bench}~\cite{hu2025repo2run} is evaluated on the 420 repositories in our three-attempt records and targets reproducible Docker environments for Python; a run succeeds when the generated Dockerfile builds and the container can run or collect pytest-based tests, such as \lt{pytest --collect-only -q}. 

\noindent\textbf{Installamatic-Bench}~\cite{milliken2025beyond} contains 40 open-source Python repository installation tasks, each with a ground-truth working installation process. The benchmark asks an agent to turn repository documentation into an executable Dockerfile and then validates the installed repository. A run succeeds if the generated Dockerfile runs the repository tests and at least one test passes, which the original benchmark treats as evidence that the test-suite dependencies were effectively installed.

\paragraph{Baselines}
For each benchmark, our primary baseline is the prior setup agent it was built for, chosen to span the dominant setup paradigms. ExecutionAgent~\cite{bouzenia2025you} performs single-command interactive setup, Repo2Run~\cite{hu2025repo2run} performs command-level setup with one-step rollback, and Installamatic~\cite{milliken2025beyond} performs documentation-guided whole-Dockerfile generation. 
We use the released ExecutionAgent, Repo2Run and Installamatic results. As additional reference points, we include reported results for general-purpose coding agents where they are available, namely
AutoGPT~\cite{yang2023auto} on ExecutionAgent-Bench,
SWE-agent~\cite{yang2024sweagent} on Repo2Run-Bench, and
OpenHands~\cite{wang2024openhands} on SetupBench.
We match HerAgent's base model to each comparison so that differences reflect harness design rather than model strength.

\paragraph{Models} To cover both modern capability and support matched-model comparisons, we evaluate HerAgent with four models: \emph{gpt-4o-mini-20240718}, \emph{gpt-4o-20241120},   \emph{gpt-4.1-20250414}, and \emph{gpt-5.4-20260305}.

\subsection{Experimental Setup}
\begin{table*}[th!]
    \centering
    \caption{Comparison of HerAgent and related setup agents across four benchmark datasets under matched-model settings.}
    \label{tab:rq1-effectiveness-efficiency}
    \begin{minipage}{0.9\textwidth}
    \footnotesize
    HerAgent reported success is any@3; baseline values follow their source protocols.\\
    \textbf{“-”} means the metric is not reported or not comparable in the original source.
    \textbf{\textsuperscript{*}} means ExecutionAgent reports per-project processing time with five projects processed in parallel.
    \textbf{$^\dagger$} means Repo2Run reports successful-build time and action usage over the 361 successfully built repositories.
    \textbf{$^\ddagger$} means Installamatic reports per-attempt installation time.
    \end{minipage}
    \footnotesize
    \setlength{\tabcolsep}{3pt}
    \renewcommand{\arraystretch}{0.95}
    \begin{tabular}{llrrrr}
    \toprule
    Dataset & Method & \makecell[c]{Success Rate(\%)} & \makecell[c]{Avg. Cost\\(\$)} & \makecell[c]{Avg. Time\\(min)} & \makecell[c]{Avg.\\Steps} \\
    \midrule
    \makecell[l]{ExecutionAgent-bench}
    & ExecutionAgent(\textit{gpt-4o-mini}) & 66.0 & 0.16 & 74\textsuperscript{*} & - \\
    & AutoGPT(\textit{gpt-4o-mini}) & 8.0 & - & - & - \\
  \rowcolor{herpurple}  & \textbf{HerAgent}(\textit{gpt-4o-mini}) & \textbf{70.0} & \textbf{0.08} & 23.1 & 30.4 \\
   \rowcolor{herpurple} & \textbf{HerAgent}(\textit{gpt-5.4}) & \textbf{80.0} & 0.18 & 13.9 & 35.4 \\
    \midrule
    \makecell[l]{Repo2Run-bench}
    & Repo2Run(\textit{gpt-4o}) & 86.0 & 3.56 & 29.0\textsuperscript{\dag} & 75.5\textsuperscript{\dag} \\
    & SWE-agent(\textit{gpt-4o}) & 9.0 & - & - & - \\
   \rowcolor{herpurple}  & \textbf{HerAgent}(\textit{gpt-4o}) & \textbf{87.1} & 0.42 & 23.9 & 51.6 \\
  \rowcolor{herpurple}   & \textbf{HerAgent}(\textit{gpt-5.4}) & \textbf{94.0} & \textbf{0.21} & 18.1 & 39.5 \\
    \midrule
    \makecell[l]{SetupBench}
    & OpenHands(\textit{gpt-4.1}) & 46.3 & 2.2 & - & 29.5 \\
    & OpenHands(\textit{Claude 4 Sonnet}) & 57.4 & 10.2 & - & 47.1 \\
 \rowcolor{herpurple}    & \textbf{HerAgent}(\textit{gpt-4.1}) & \textbf{70.4} & \textbf{0.12} & 9.11 & 27.5 \\
 \rowcolor{herpurple}    & \textbf{HerAgent}(\textit{gpt-5.4}) & \textbf{90.7} & 0.30 & 21.2 & 51.3 \\
    \midrule
    \makecell[l]{Installamatic-bench}
    & Installamatic(\textit{gpt-4o-mini}) & 52.0 & - & 8.35\textsuperscript{\ddag} & - \\
 \rowcolor{herpurple}    & \textbf{HerAgent}(\textit{gpt-4o-mini}) & \textbf{85.0} & \textbf{0.02} & 7.79 & 35.6 \\
   \rowcolor{herpurple}  & \textbf{HerAgent}(\textit{gpt-5.4}) & \textbf{92.5} & 0.45 & 20.65 & 54.7 \\
    \bottomrule
    \end{tabular}
    \end{table*}

HerAgent is implemented in Python and completes setup tasks inside isolated Docker containers. All LLM calls are made through the OpenAI API with temperature 0.7 and at most three request attempts. We use the models reported in each experiment. Each run starts from an \texttt{ubuntu:24.04} base image with build tools, Git, \texttt{curl}, \texttt{wget}, bash-compatible shell settings, and the Docker CLI. Command execution and Docker image builds both use a 1800-second timeout.\footnote{The complete generated Dockerfiles and run artifacts are available in \artifactrepo.} The Docker runtime is Docker Engine Community 29.4.2 on both client and server, Docker API 1.54, containerd 1.7.28, runc 1.3.0, and Linux kernel 6.8.0-110-generic.

\paragraph{Methodology for RQ1}
For overall effectiveness, each evaluated task starts from a clean environment and uses a fixed repository version, base image, and budget within each dataset. To control for model strength, HerAgent and the corresponding baseline on each dataset use the same base model. 

Table~\ref{tab:rq1-effectiveness-efficiency} reports success and efficiency metrics.
\emph{Success Rate} is the percentage of tasks whose final environment passes the dataset-provided validation target. {Success uses \emph{any@3}: a task succeeds if at least one of three independent runs passes the dataset-provided external oracle. Thus, an experiment with $N$ tasks comprises $3N$ independent setup attempts; these runs are distinct from API request retries. Table~\ref{tab:three-run-distribution} reports the per-task success counts and each run's success rate. Efficiency metrics are aggregated over the three runs per task. ExecutionAgent uses any@3, Installamatic uses any@10, and Repo2Run uses any@3. These aggregation differences should be considered when interpreting the released baseline results.}
\emph{Average Cost} is the mean LLM API cost per task in U.S. dollars, computed from input and output token usage using the corresponding model prices.\footnote{\url{https://platform.openai.com/docs/pricing}, accessed June 25, 2026.} 
For HerAgent, \emph{Average Time} is the mean execution time per task in minutes, including all phases. 
\emph{Average Steps} is the mean number of agent-environment interaction steps. One step is one action that submits an executable setup action to the sandbox and receives an observation. For command-level agents, this is typically one command execution; for HerAgent, it includes a block execution, validation, repair, rollback, and replay action. Thus, steps measure interaction count rather than the number of shell lines in the final setup script.

\begin{table*}[tb]
\centering
\caption{Supplementary external-oracle results from three HerAgent attempts per repository.}
\label{tab:three-run-distribution}
\begin{minipage}{0.65\textwidth}
\footnotesize
0/3--3/3 count repositories by successful attempts.\\
R1--R3 are round-wise success rates (\%).
\end{minipage}
\footnotesize
\setlength{\tabcolsep}{4pt}
\begin{tabular}{ll|rrrr|rrr}
\toprule
Benchmark & Model & 0/3 & 1/3 & 2/3 & 3/3 & R1 (\%) & R2 (\%) & R3 (\%) \\
\midrule
\rowcolor{blue!5} ExecutionAgent & GPT-4o-mini & 15 & 1 & 12 & 22 & 66.0 & 52.0 & 64.0 \\
\rowcolor{blue!5}  & GPT-5.4 & 10 & 2 & 15 & 23 & 68.0 & 64.0 & 70.0 \\
Repo2Run & GPT-4o & 54 & 38 & 133 & 195 & 56.9 & 76.7 & 78.1 \\
 & GPT-5.4 & 25 & 10 & 128 & 257 & 75.2 & 86.0 & 85.7 \\
\rowcolor{blue!5} SetupBench & GPT-4.1 & 16 & 3 & 13 & 22 & 48.1 & 63.0 & 64.8 \\
\rowcolor{blue!5}  & GPT-5.4 & 5 & 5 & 19 & 25 & 68.5 & 75.9 & 74.1 \\
Installamatic & GPT-4o-mini & 6 & 5 & 9 & 20 & 55.0 & 75.0 & 77.5 \\
 & GPT-5.4 & 3 & 0 & 15 & 22 & 75.0 & 75.0 & 90.0 \\
\bottomrule
\end{tabular}
\end{table*}
Across all HerAgent experiments, we limit local repair attempts to 20 total per task. This global budget includes repairs attempted within the current block and repairs attempted after rollback to an earlier block; together, they represent the agent's overall allowance for correcting the current failed setup state during one setup attempt. If the budget is exceeded, HerAgent stops the current setup attempt and marks the run as failed.

\paragraph{Methodology for RQ2} We run the progress-control ablation on SetupBench (Repo Setup subset){ , chosen for its diverse setup tasks across seven language ecosystems.}
All RQ2 variants use \emph{gpt-5.4-20260305} and follow the RQ1 setup, including the same token budgets, and attempt budget.
For fairness, command-level variants run in a persistent shell managed by \texttt{pexpect}, preserving ordinary shell state such as the working directory and exported variables. 

Because the progress-control mechanisms are coupled, 
rather than removing one component at a time, we vary along two axes, namely the unit of forward execution
and the recovery policy (Table~\ref{tab:rq2-progress}).
\emph{(1) w/o Semantic Boundary} keeps block-level execution and recovery as full HerAgent, but mechanically splits the setup script into eight fixed command chunks instead of semantic blocks;
\emph{(2) w/o Patch-Back} keeps block-level forward execution and live repair, but does not write successful live repairs back into the replayable setup artifact;
\emph{(3) w/o Local Repair} keeps block-level forward execution but replaces localized repair with regeneration of the failed block;
{\emph{(4) w/o Block-Level Forward} replaces block-level forward execution with command-level execution and uses live repair with patch-back;}
\emph{(5) w/o Block-Level Recovery} also uses command-level execution, but recovers by command rollback rather than checkpointed block-level repair;
\emph{(6) w/o Block Partitioning} removes block boundaries entirely and falls back to whole-script regeneration after failures.

Full HerAgent is the reference configuration.
\begin{table*}[tb]
    \centering
    \caption{RQ2 Progress-Control Component Ablations on SetupBench}
    \label{tab:rq2-progress}
    \footnotesize
    \setlength{\tabcolsep}{3.0pt}
    \renewcommand{\arraystretch}{1.0}
    \resizebox{\textwidth}{!}{%
    \begin{tabular}{@{}p{0.25\textwidth}|p{0.12\textwidth}|p{0.19\textwidth}|rrrrrr@{}}
    \toprule
    Method & Forward Mode & Recovery Mode &
    \makecell[c]{Forward\\Succ. $\uparrow$} &
    \makecell[c]{Replay\\Succ. $\uparrow$} &
    \makecell[c]{Final\\Succ. $\uparrow$} &
    \makecell[c]{Avg.\\Cost $\downarrow$} &
    \makecell[c]{Avg.\\Time $\downarrow$} &
    \makecell[c]{Avg.\\Steps $\downarrow$} \\
    \midrule
    \textbf{HerAgent}
    & \textbf{block} & \textbf{checkpointed local repair}
    & \textbf{96.3} & \textbf{96.3} & \textbf{90.7} & 0.308 & 21.2 & 51.3 \\
    \midrule
    \hspace{1em}-w/o Semantic Boundary
    & \multirow{4}{*}{block} & checkpointed local repair
    & 72.2 & 68.5 & 48.1 & 0.326 & 19.7 & 30.3 \\
    \hspace{1em}-w/o Patch-Back
    & & live repair w/o patch-back
    & 66.7 & 37.0 & 25.9 & 0.074 & 6.19 & 27.2 \\
    \hspace{1em}-w/o Local Repair
    & & regenerate failed block
    & 96.3 & 90.7 & 74.1 & 0.120 & 7.94 & 32.0 \\
    \midrule
    \hspace{1em}{-w/o Block-Level Forward}
    & \multirow{2}{*}{command} & {live repair with patch-back}
    & {88.9} & {87.0} & {77.8} & {0.168} & {5.26} & {53.4} \\
    \makecell[l]{\hspace{1em}-w/o Block-Level Recovery}
    & & command rollback
    & 70.4 & 68.5 & 53.7 & 0.176 & 3.61 & 58.7 \\
    \midrule
    \hspace{1em}-w/o Block Partitioning
    & whole script & regenerate full script
    & 55.6 & 53.7 & 46.3 & 0.372 & 15.9 & 28.8 \\
    \bottomrule
    \end{tabular}}
    \end{table*}
We report three stage-wise success metrics, aligned with the workflow in Figure~\ref{fig:approach}. 
\emph{Forward Succ.} means that all setup blocks or generated setup units have executed and passed their internal validation before final replay. 
{ \emph{Replay Succ.} is the percentage of all tasks whose final artifact replays successfully from a clean environment and passes HerAgent's self-generated block checks $V$. These intermediate checks are distinct from the external benchmark oracle.}
\emph{Final Succ.} means that the replayed environment also passes the dataset-provided validation oracle, matching the \emph{Success Rate} metric in Table~\ref{tab:rq1-effectiveness-efficiency}. 
Cost, time, and steps also follow the RQ1 metric definitions.

\paragraph{Methodology for RQ3 and RQ4.} RQ3 and RQ4 analyze the HerAgent trajectories reported in Table~\ref{tab:rq1-effectiveness-efficiency}.
For RQ3, we use the seven block-template categories in Table~\ref{tab:block-templates} as functional labels; the actual generated sequence may merge, omit, or add project-specific blocks. Using these categories, we measure block frequency and commands per block.
We then label trajectory events by progress-control action type. As illustrated in Figure~\ref{fig:rq3-case}, forward actions advance along the planned block sequence (steps~\#1--\#2 and~\#4--\#5), local-repair actions patch a failed block from observed errors (step~\#3), and rollback actions restore an earlier clean checkpoint after root-cause relocation (steps~\#6--\#9).
Using these labels, we measure repair and rollback locations, rollback outcomes, and repair counts by final outcome. Cost/time shares are computed within the trajectories containing each action type; e.g., rollback share is measured only over projects that invoke rollback.

{For RQ4, we checked 1,019 failure records in RQ1. We exclude records caused by external conditions, such as API or network failures, unavailable credentials, insufficient disk space, or missing host hardware. The final remaining records number is 904.}
{The annotation procedure follows three stages.
First, an LLM (we used gpt-5.4-20260305 here) summarized the failure evidence, suggested a possible failure category, and drafted initial root-cause label; these outputs were not used as final labels. 
Second, two researchers reviewed the definitions, agreed on a shared codebook, and independently examined each case, including its trajectory, logs, generated artifacts, and validation errors, to assign one primary root-cause label. 
Third, after the discussion and adjudication, the researchers re-examined disagreements and revisited affected labels.}

\section{Experimental Results}
\label{sec:results}

In this section we answer our four research questions 
and discuss potential threats to the validity of our study.


\subsection{Results for RQ1: Overall Effectiveness}
\label{sec:rq1}
Table~\ref{tab:rq1-effectiveness-efficiency} compares HerAgent with prior setup agents across four benchmarks, reporting success rate and efficiency metrics under matched-model settings plus an additional stronger-model HerAgent run.
The columns report effectiveness (success rate) together with efficiency metrics. The shaded rows highlight HerAgent for easier comparison.

Regarding \emph{effectiveness}, across all four benchmarks, HerAgent achieves the highest success rate over the matched baselines: 66.0\% to 70.0\% against ExecutionAgent on ExecutionAgent-Bench (+4.0 percentage points), 86.0\% to 87.1\% against Repo2Run on Repo2Run-Bench (+1.1 percentage points), 46.3\% to 70.4\% against OpenHands on SetupBench  (+24.1 percentage points), and 52.0\% to 85.0\% against Installamatic on Installamatic-Bench  (+33.0 percentage points). General code agents remain much weaker in this setting, with AutoGPT at 8.0\%, SWE-agent at 9.0\%, and OpenHands at 46.3\% on their corresponding datasets. With \emph{gpt-5.4}, HerAgent further reaches 80.0\%, 94.0\%, 90.7\%, and 92.5\% on the four datasets.

Regarding \emph{efficiency}, HerAgent also reduces cost in the matched-model settings. On SetupBench, it costs \$0.12 per task versus \$2.20 for OpenHands, a 94.5\% reduction. On Repo2Run-Bench, it costs \$0.42 versus \$3.56 { (88.2\% less)} and reduces average time from 29.0 to 23.9 minutes. On ExecutionAgent-Bench, it costs \$0.08 versus \$0.16 and completes in 23.1 minutes, compared with 74 minutes for ExecutionAgent.

\begin{tcolorbox}[colback=lightgray!20, colframe=lightgray!50, arc=2pt, boxrule=0.5pt]
{\small
\textbf{Answer to RQ1:} {HerAgent improves setup success by 33.0 percentage points over Installamatic on Installamatic-Bench with GPT-4o-mini, and reduces API cost by 94.5\% relative to OpenHands on SetupBench with GPT-4.1.}
}
\end{tcolorbox}

\subsection{Results for RQ2: Component Contribution}
\label{sec:rq2}

\begin{figure*}
\centering
\includegraphics[width=0.9\linewidth]{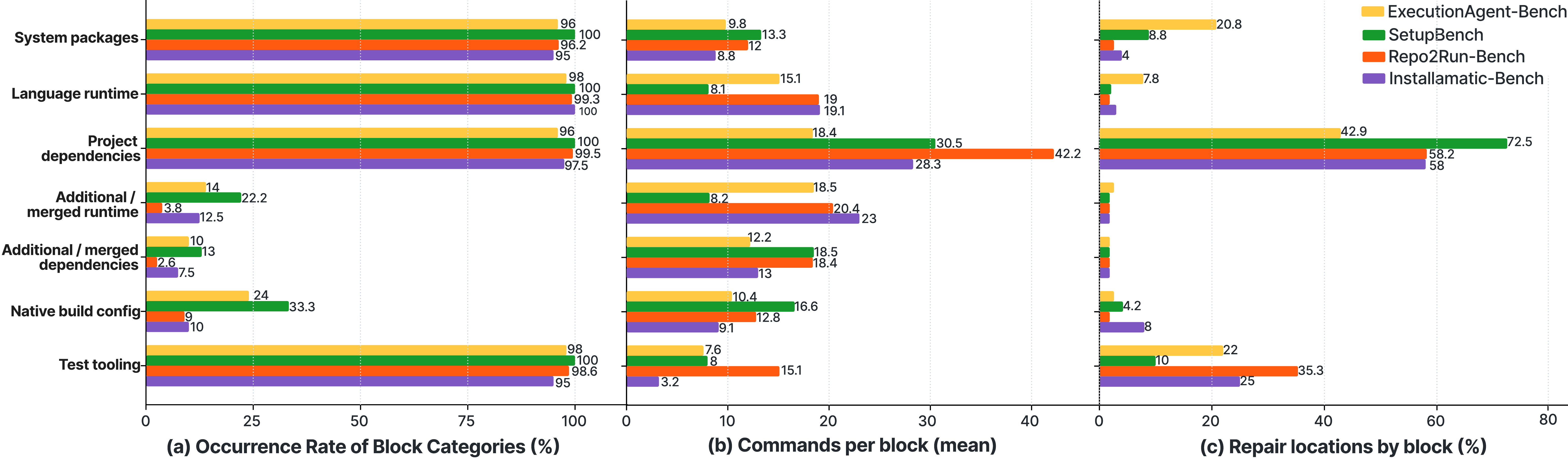}
\caption{Block-level setup structure across benchmarks. Panel (a) reports the occurrence frequency of each block category. {Panel (b) reports the mean number of commands per block.} Panel (c) reports the share of local repairs located in each block category.}
\label{fig:block_statistics}
\end{figure*}

Table~\ref{tab:rq2-progress} reports RQ2 ablation results. Full HerAgent incorporates the full combination of block-level progress, checkpointed repair, patch-back, and replay validation. It performs best across all stages: 96.3\% forward success, 96.3\% replay success, and 90.7\% final success.

We organize the variants by their
forward-execution unit: the block-level variants hold forward execution fixed and vary recovery; {the command-level variants assess command-level execution with different recovery policies.}  A final variant removes block
boundaries entirely.

\vspace{0.4ex}
\noindent\textbf{Block-level forward, varying recovery.}
\emph{w/o Semantic Boundary} keeps HerAgent's block-level execution and recovery, but replaces semantic blocks with eight fixed command chunks. Its forward success drops to 72.2\%, replay success to 68.5\%, and final success to 48.1\%. This shows that block-level control alone is not enough: the boundaries must align with setup functions such as dependency resolution and validation preparation.
\emph{w/o Local Repair} replaces in-block repair with whole-block regeneration while keeping block boundaries and checkpointing. It matches Full HerAgent on forward success (96.3\%),
since regenerated blocks still pass on the way through. However, final success falls to 74.1\%. This suggests that a failed block often contains correct logic that should be retained by focusing specifically on addressing the failure symptom within a block (rather than regenerating a block from scratch).
\emph{w/o Patch-Back} repairs failed blocks but never writes the fix into the output artifact. The repaired containers pass, but final success drops to 25.9\%, because the repairs are not saved for clean replay. Live repair without patch-back is therefore largely non-reproducible. 

\noindent\textbf{Command-level forward, varying recovery.} 
{\emph{w/o Block-Level Forward} uses command-level execution and live repair with patch-back, reaching 88.9\% forward success, 87.0\% replay success, and 77.8\% final success. Replay success is close to forward success, but final success remains 12.9 percentage points below Full HerAgent (90.7\%). This result suggests that command-level execution with patch-back supports replayability, while block-level progress control provides additional benefits for external-oracle success.}
{\emph{w/o Block-Level Recovery} uses command rollback and reaches 70.4\% forward success, 68.5\% replay success, and 53.7\% final success. It performs worse than the command-level variant with live repair and patch-back across all three stages. The comparison shows that these command-level recovery policies differ substantially in effectiveness; command rollback alone does not recover the success rate of Full HerAgent.}

\noindent\textbf{Block boundaries quality and absence.}
The results of \emph{w/o Semantic Boundary} show that block-level control is insufficient when boundaries are arbitrary fixed chunks rather than semantically aligned units.
\emph{w/o Block Partitioning} removes block boundaries entirely and reaches only 46.3\% final success, showing that whole-script regeneration is too coarse. 
These variants show that boundaries matter in two ways: HerAgent needs blocks, and those blocks need to align with setup functions. 

\noindent\textbf{Efficiency.} 
The efficiency results show that HerAgent's higher cost buys reliability rather than wasted work. 
{Removing patch-back reduces cost but also final success:} \emph{w/o Patch-Back} is cheapest (\$0.074, 6.19 minutes, 27.2 steps) but reaches only 25.9\% final success.
Comparable spending is also insufficient without semantic boundaries: \emph{w/o Semantic Boundary} costs \$0.326 and takes 19.7 minutes, close to Full HerAgent, but reaches only 48.1\% final success. 
{\emph{w/o Block-Level Forward} costs \$0.168 and takes 5.26 minutes and 53.4 steps, achieving 77.8\% final success at lower cost and time than Full HerAgent. The command-level variants require 53.4 and 58.7 steps, respectively, compared with 51.3 for Full HerAgent.} By comparison, \emph{w/o Block Partitioning} has fewer steps (28.8) but the highest cost (\$0.372), showing that whole-script regeneration makes each attempt expensive.


\begin{tcolorbox}[colback=lightgray!20, colframe=lightgray!50, arc=2pt, boxrule=0.5pt]
{\small
\textbf{Answer to RQ2.}
The ablation shows the value of all HerAgent design choices. Semantic block boundaries are essential, not merely command grouping. Block-level execution prevents fragmented progress, checkpointed recovery improves replayability, local repair avoids destructive regeneration, and patch-back makes fixes reproducible. 
}
\end{tcolorbox}

\subsection{Results for RQ3: Trajectory-Level Recovery Analysis}
\label{sec:rq3}

\begin{figure}[tb]
\centering
\includegraphics[width=0.9\linewidth]{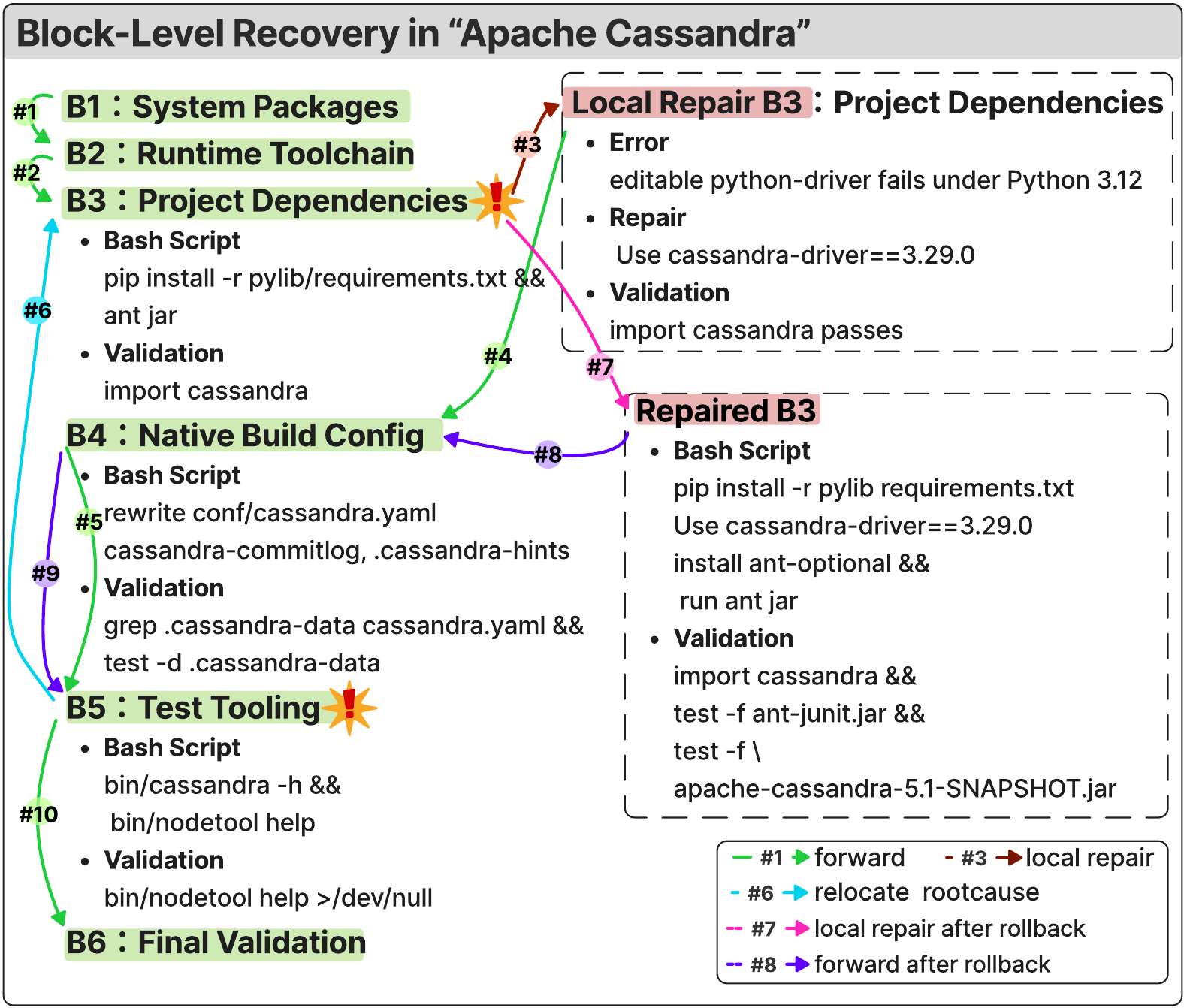}
\caption{Apache Cassandra recovery trace. The failure is exposed in the later test-tooling block ($B_5$), but HerAgent relocates the root cause to the earlier dependency/build-preparation block ($B_3$), rolls back to checkpoint \emph{Ckpt2}, patches the repaired $B_3$, and replays the downstream suffix.}
\label{fig:rq3-case}
\end{figure}

\begin{figure*}[tb]
\centering
\includegraphics[width=\textwidth]{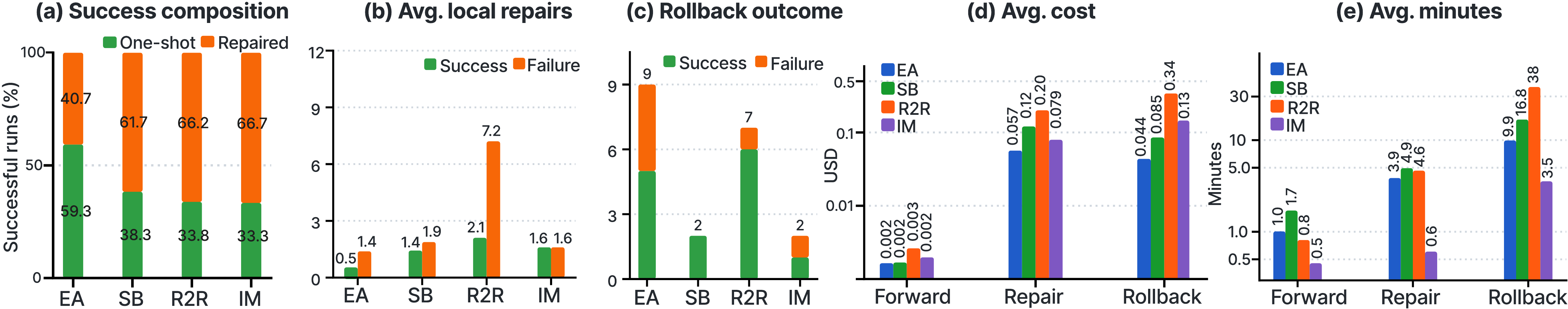}
\caption{Trajectory-level recovery behavior. (a) One-shot vs. repair-assisted final successes. (b) Average number of local repairs for successful and failed projects in each benchmark. (c) Outcomes after root-cause relocation and rollback. (d,e) Average cost and time by progress-control action type.\\
Abbrev.: EA=ExecutionAgent-Bench, SB=SetupBench, R2R=Repo2Run-Bench, IM=Installamatic-Bench.}
\label{fig:rq3-traj}
\end{figure*}

\paragraph{Block structure}
Figure~\ref{fig:block_statistics}(a) shows that HerAgent's generated scripts show common setup patterns. \emph{System packages}, \emph{Language runtime}, \emph{Project dependencies}, and \emph{Test tooling} appear in almost every run across the four datasets. The ``additional'' block categories are more selective: \emph{Additional / merged runtime} and \emph{Additional / merged dependencies} are less common. \emph{Native build config} is most common on SetupBench (33.3\%) but much less common on Repo2Run-Bench (9.0\%) and Installamatic-Bench (10.0\%).

Figures~\ref{fig:block_statistics}(b) and~\ref{fig:block_statistics}(c) show that setup and recovery workloads concentrate in the same two stages.  \emph{Project dependencies} is the largest stage (18.4--42.2 commands across datasets). It is also a major repair location because dependency and version choices introduce many downstream failures. \emph{Test tooling} is usually smaller, but is also a frequent repair site.  
Together, \emph{Project dependencies} and \emph{Test tooling} account for 64.9\%--93.5\% of local repairs, showing that most repair effort falls on dependency resolution and validation preparation. 
HerAgent's block boundaries therefore delineate stages that dominate setup, dependency resolution and validation, which is where checkpoints and recovery are most useful.

\paragraph{Trajectory analysis}
Figure~\ref{fig:rq3-traj} characterizes where recovery is needed and where recovery cost is spent. Panel (a) shows that many final successes are not one-shot successes but require repair, indicating that environment setup frequently needs iterative recovery rather than a single generated setup script. { Among successful runs, 59.3\%, 38.3\%, 33.8\%, and 33.3\% require no repair on ExecutionAgent-Bench, SetupBench, Repo2Run-Bench, and Installamatic-Bench, respectively. These are conditional shares among successes, rather than standalone single-shot success rates over all tasks.
For context, the prior studies report single-shot versus full-system success of 10.5\% versus 28.8\% for Installamatic~\cite{milliken2025beyond}, 10.0\% versus 66.0\% for ExecutionAgent~\cite{bouzenia2025you}, and 22.1\% versus 86.0\% for Repo2Run~\cite{hu2025repo2run}. These within-study comparisons support iterative recovery; their original settings, particularly Installamatic's 28.8\% full-system result, should not be conflated with the released baseline setting in Table~\ref{tab:rq1-effectiveness-efficiency}.}
Panel (b) compares local-repair intensity between successful and failed projects. Failed runs generally require more repairs, but the gap varies by benchmark: it is small on SetupBench (1.4 vs. 1.9 repairs per project) and absent on Installamatic-Bench (1.6 vs. 1.6), larger on ExecutionAgent-Bench (0.5 vs. 1.4), and largest on Repo2Run-Bench (2.1 vs. 7.2). These results show both the value and limit of local repair: it handles routine block-level faults, but repeated repairs without convergence signal that the failure likely requires broader recovery than patching the current block.
Panel (c) isolates the less frequent but more consequential rollback cases. Although root-cause relocation and rollback are triggered only in a subset of trajectories, they often turn otherwise stuck runs into successes: 5/9 rollback cases succeed on ExecutionAgent-Bench, 2/2 on SetupBench, 6/7 on Repo2Run-Bench, and 1/2 on Installamatic-Bench. This shows that rollback is not a routine repair step, but a targeted recovery mechanism for failures whose cause lies before the observed failing block.

Panels (d) and (e) report the average cost and time for each progress-control action type. Forward execution runs the next planned block and is cheap (\$0.002--\$0.003; 0.5--1.7 minutes). Local repair diagnoses a failed block, tries candidate fixes, and validates the patched block, making it costlier (\$0.057--\$0.20; 0.6--4.9 minutes). Rollback is the most expensive in wall-clock time because it follows root-cause relocation, restores an earlier checkpoint, and replays downstream blocks, reaching 9.9, 16.8, and 38.0 minutes on three benchmarks. This supports HerAgent's policy of using local repair first and reserving rollback for non-local faults.

\paragraph{Qualitative recovery case}
Figure~\ref{fig:rq3-case} shows a SetupBench “Apache Cassandra” case with two recovery actions. The target setup starts a single-node Cassandra instance and validates it with \texttt{bin/nodetool status}. 
The first failure occurs while executing $B_3$, where HerAgent applies local repair directly to the failed block(arrow \#3 in Figure~\ref{fig:rq3-case}). After this repair is patched back and replayed, HerAgent continues forward(arrow \#4 and \#5).
The second failure is exposed later in the test-tooling block $B_5$, where \texttt{bin/nodetool help} fails. Instead of repeatedly patching $B_5$, HerAgent relocates the root cause to the earlier $B_3$ block(arrow \#6). It rolls back to \emph{Ckpt2}, repairs $B_3$ by replacing the problematic editable \texttt{datastax/python-driver} installation with \texttt{cassandra-driver==3.29.0}, adding missing helper dependencies, and ensuring the required Cassandra build artifacts are available(arrow \#7). HerAgent then patches back the repaired $B_3$, replays $B_4$(arrow \#8) and $B_5$(arrow \#9), and finally advances to $B_6$(arrow \#10).

The same case exposes the failure modes of the RQ2 ablations(Section \ref{sec:rq2}). Command-level variants tend to repair the observed \texttt{nodetool} symptom in $B_5$ while leaving the earlier dependency/build state in $B_3$ unresolved. The whole-script variant has the opposite problem: it can change $B_3$ only by regenerating a much larger script, risking unrelated rewrites and lost progress. This shows why block-level rollback is useful: it repairs the responsible stage without discarding the validated prefix.

\begin{tcolorbox}[colback=lightgray!20, colframe=lightgray!50, arc=2pt, boxrule=0.5pt]
{\small
\textbf{Answer to RQ3.}
HerAgent's block structure is stable and recovery-relevant: dependency and test-tooling stages dominate both workload and repairs. Local repair is effective for common block-level faults. Rollback is less frequent but crucial when the root cause lies in an earlier block. 
}
\end{tcolorbox}

\subsection{Results for RQ4: Failure Analysis}
\label{sec:rq4}
\begin{figure}[t]
\centering
\includegraphics[width=0.6121\linewidth]{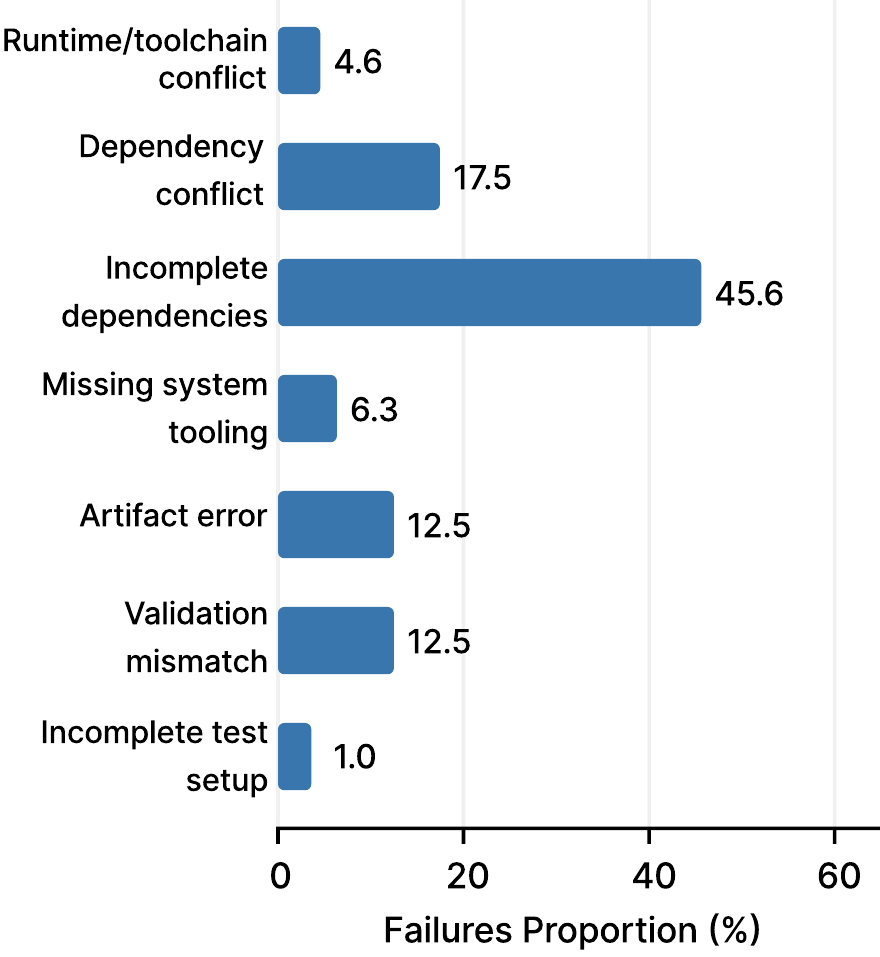}
\caption{{Setup failure reasons across the classified trajectory records remaining after external-condition exclusions.}}
\label{fig:rq4-failure-reasons}
\end{figure}

{
During the independent human annotation stage, the two researchers' initial labels achieved an observed agreement of 69.96\% and an expected agreement by chance of 24.66\%, yielding Cohen's $\kappa=0.601$. After disagreements were resolved, the final labels were used to produce the failure-category distribution in Figure~\ref{fig:rq4-failure-reasons} for the records retained after the external-condition exclusions described in the methodology. Incomplete dependencies account for 45.6\%, dependency conflicts for 17.5\%, artifact errors and validation mismatch for 12.5\% each, missing system tooling for 6.3\%, runtime/toolchain incompatibility for 4.6\%, and incomplete test setup for 1.0\%. Together, the two dependency categories account for 63.1\%. We present one representative example for each category below.

\paragraph{Runtime/toolchain conflict}
The selected language runtime or build tool does not meet the project's version requirements. For example, \texttt{denoland/deno} requires \texttt{edition2024}, which Cargo 1.75.0 cannot support. Package-level API/ABI incompatibilities are classified separately as dependency conflicts.

\paragraph{Dependency conflict}
Package constraints cannot be satisfied together, or installed packages expose incompatible APIs or ABIs. In \texttt{pi-nexus-autonomous-banking-network}, conflicting Flask requirements, including \texttt{Flask==2.0.1}, cause pip to terminate with \texttt{ResolutionImpossible}. 

\paragraph{Incomplete dependencies}
Required packages, build backends, local dependencies, or distribution files are absent or unavailable. In \texttt{pandas-dev/pandas}, the build cannot import \texttt{mesonpy}; we classify this as an incomplete dependency declaration or installation.

\paragraph{Missing system tooling}
A required executable, system development package, or native library is absent. In \texttt{Auralis}, the Python package \texttt{sounddevice} is installed, but importing it raises \texttt{OSError: PortAudio library not found}. 

\paragraph{Artifact error}
Generated setup or repair artifacts contain syntax errors, incorrect paths or working directories, or omit files they should create. In \texttt{ansible/ansible}, generated inline Python contains \texttt{. = Path(sys.argv[1])}, causing \texttt{SyntaxError: invalid syntax} before the intended operation can execute.

\paragraph{Validation mismatch}
The validation entry point, discovery scope, arguments, or interpreter does not match the project. In \texttt{cookiecutter-mlops-package}, pytest collects an unrendered template containing a Cookiecutter placeholder in a Python import, causing a syntax error. This is a validation-entry mismatch; changing a pytest exit code cannot substitute for discovering and running the intended tests.

\paragraph{Incomplete test setup}
An applicable validation procedure lacks required project-local initialization, test data, directories, or launch context. In \texttt{kronfluence}, collection of distributed tests raises \texttt{KeyError: 'LOCAL\_RANK'}, so the record is classified as incomplete test setup. 
}

\begin{tcolorbox}[colback=lightgray!20, colframe=lightgray!50, arc=2pt, boxrule=0.5pt]
{\small
\textbf{Answer to RQ4.}
Dependency conflicts and incomplete dependencies account for 63.1\% of the classified records. The seven categories distinguish compatibility and installation problems from system prerequisites, artifact defects, validation selection, and test initialization. They motivate compatibility-aware dependency handling, executable artifact checks, and project-appropriate validation with explicit prerequisites.
}
\end{tcolorbox}

\subsection{Threats to Validity}
\label{sec:threats}

The main threats concern benchmark scope, baseline reproducibility, and validation strength. The four datasets may not represent environments that require private credentials, paid services, GPUs, or complex distributed deployment. Some benchmark do not release complete prompts, commits, model versions, or validation commands, so comparisons may still reflect repository-version drift or differences in available execution details. Network availability and package versions can also change over time; we mitigate this by recording timestamps, base images, logs, commits, and generated artifacts. Finally, a dataset-specific readiness command does not prove full repository usability, and LLM-based setup agents can vary across runs. We therefore interpret success only against each benchmark oracle and avoid overclaiming beyond the evaluated settings.

\section{Related Work}

{
\noindent\textbf{Environment construction and reproducible artifacts.}
Executable environments are a prerequisite for the test feedback used by SWE-bench~\cite{jimenez2024swebench} and engineering agents such as SWE-agent~\cite{yang2024sweagent} and OpenHands~\cite{wang2024openhands}.
Environment-specific benchmarks make this prerequisite an explicit evaluation target: SetupBench~\cite{arora2025setupbench} and EnvBench~\cite{eliseeva2025envbench} evaluate environment bootstrapping, while ResearchEnvBench~\cite{wang2026researchenvbench} and CSR-Bench~\cite{xiao2025csrbench} examine the execution and deployment of research repositories.
RAT~\cite{huang2026rat} and Multi-Docker-Eval~\cite{fu2025multidockereval} further study automated environment configuration and building.
These efforts make dependency, toolchain, service, and validation requirements explicit parts of repository execution.
At the system level, MEnvAgent~\cite{guo2026menvagent} addresses scalable polyglot construction, EvoConfig~\cite{guo2026evoconfig} studies self-evolving multi-agent configuration, and DockSmith~\cite{zhang2026docksmith} builds coding environments through an agentic Docker builder.
SWE-Factory~\cite{guo2025swefactory}, daVinci-Env~\cite{fu2026davincienv}, and SWE-Universe~\cite{chen2026sweuniverse} connect environment construction to the production of verifiable engineering tasks and training data.
Reproducibility also has an established systems foundation: Docker~\cite{merkel2014docker} provides container isolation, Nix~\cite{dolstra2004nix} specifies software deployment, and ReproZip~\cite{chirigati2016reprozip} captures dependencies for reproducing executions.
HerAgent complements these efforts by organizing partial setup progress into functional blocks that link validated checkpoints and recorded repairs to a reconstructible artifact.

\noindent\textbf{Execution granularity, validation, and failure recovery.}
Existing setup agents provide complementary ways to organize interaction and repair.
ExecutionAgent~\cite{bouzenia2025you} executes individual commands, and Repo2Run~\cite{hu2025repo2run} augments command-level environment construction with snapshots and rollback to the state preceding a failed command.
PiPer~\cite{kovrigin2025piper} and Installamatic~\cite{milliken2025beyond} organize setup around generated scripts or Dockerfiles, while Vibenix~\cite{schwaighofervibenix} targets software packaging with Nix.
SetupX~\cite{zhou2026setupx} combines reusable failure experience with speculative execution and snapshot-based rollback to earlier execution states.
Command-level rollback can revisit earlier commands, but its recovery boundaries do not explicitly group commands by setup function.
HerAgent instead rolls back and repairs a complete functional block, so it can correct an earlier setup stage while preserving the validated prefix.
Accepted fixes are patched back into the setup artifact and checked by clean replay.
}

\section{Conclusions}
\label{sec:conclusions}

HerAgent treats environment setup as a progress-control problem and advances it over validated functional blocks, repairing and rolling back at block boundaries so that fixes become reproducible setup artifacts.
Our ablation results substantiate the role of block-level design. Without such boundaries, a method may either roll back too locally and keep polluted state, or reset too globally and discard useful progress. The checkpointed block sequence gives the controller a middle granularity that is large enough to make progress but small enough to repair.
Additionally, 
HerAgent importantly distinguishes the current live container, the validated setup prefix, and the final replayable artifact; this is an important design improvement over prior agents, which often conflate them. This is a problem because 
setup trajectories regularly mix 
routine progress, late-exposed failures, hidden state pollution, and repairs, that must ultimately become reproducible artifacts.
The remaining failures in our experiments point beyond progress control to global compatibility reasoning, validation-command interpretation, and separating repository faults from harness faults. We see block-level checkpointing as a foundation that these capabilities can build on.

\section{Data Availability}

Our code and data are available at \artifactrepo.

\bibliographystyle{IEEEtran}
\newcommand{\BIBdecl}{\interlinepenalty=10000}
\bibliography{IEEEexample}

\end{document}